\documentclass[twoside]{article}

\usepackage{graphicx}

\usepackage{cmpj2e}

\title[Emptiness formation probability in XX chains]%
{The emptiness formation probability correlation in homogeneous and dimerized XX chains}
\author[Stolze and Garske]{J. Stolze\refaddr{label1},
        T. Garske\refaddr{label1, label2}}
\addresses{
\addr{label1} Institut f\"ur Physik, Technische Universit\"at Dortmund, 44221 Dortmund, Germany
\addr{label2} MRC Centre for Outbreak Analysis and Modelling, Department of Infectious Disease Epidemiology, Imperial College London, Norfolk Place, London W2 1PG, UK (present address).
}
\begin{document}

\maketitle

\bibliographystyle{prsty}

\begin{abstract}
We review some known properties of the ``emptiness formation
probability'' correlation  which we have calculated
numerically for spin-1/2 XX chains with constant (homogeneous) or alternating (dimerized) nearest-neighbor coupling  and an external field (in $z$
direction) for arbitrary temperature. The long-distance asymptotic behavior
of this correlation is known to be Gaussian at zero temperature and exponential at finite temperature for the homogeneous chain. By simple analytical arguments the exponential behavior at finite temperature extends to the dimerized system. Numerical results for the dimerized chain confirm the exponential decay at finite temperature and show Gaussian decay at zero temperature.
\end{abstract}

\section{Basic properties of the EFP correlation}
\label{I}

The ``emptiness formation probability'' (EFP) $P(n)$ was first \cite{KIEU94} defined as the
probability of finding a ``ferromagnetic island'' of  $n$ sites in the
ground state (or equilibrium state at finite temperature)
 of a spin-$\frac{1}{2}$ antiferromagnetic chain. For
obvious reasons it has also \cite{EFIK95b} been called ``ferromagnetic
string formation probability''. From an experimental point of view
$P(n)$ is a complicated many-particle correlation. However,
mathematically speaking, the EFP can be introduced as a natural
elementary building block for constructing various correlation
functions of integrable spin chains by using Bethe Ansatz techniques
\cite{KBI93}, see also comments given in \cite{KMST02b}. Depending on the situation at hand, a variety of
analytical techniques was applied to derive more or less explicit
results for the EFP or its long-distance asymptotics. The EFP can be
represented as a multiple integral \cite{KMST02,KMST02b}, as a
Fredholm determinant \cite{KIEU94} or as a 
Toeplitz determinant \cite{STN01,Fra08}.
 Standard numerical techniques for one-dimensional quantum
systems were used to obtain numerical data to which analytic or
asymptotic results could be compared, for example the density matrix
renormalization group (DMRG) at temperature $T=0$ \cite{STN01} or
quantum Monte Carlo simulations (QMC) at finite $T$ \cite{BKNS02}.

One model for which the EFP was discussed 
 is the nearest-neighbor $S=\frac{1}{2}$ $XXZ$
antiferromagnetic chain in an external field:
%\begin{widetext}
\begin{equation}
  H= J \sum_{i=1}^N 
  \left(
S_i^x S_{i+1}^x +
S_i^y S_{i+1}^y +
\Delta \left( S_i^z S_{i+1}^z -\frac{1}{4} \right)
\right)
-h \sum_{i=1}^N S_i^z \label{eq:1.1}.
\end{equation}
For the special case $\Delta=0$ (in which the model is known as the XX chain)  a
Gaussian decay of $P(n)$ at long distances $n$ was derived
\cite{EFIK95b} for $T=0$, and it was conjectured that the behavior is
Gaussian also for nonzero $\Delta$. Subleading correction terms to the
$\Delta=0$ asymptotic behavior were found in \cite{STN01}; the present
knowledge is summarized in the formula:
\begin{equation}
  \label{eq:1.2}
  \ln P(n) = \frac{n^2}{2} \ln \left(  \frac{1}{2J} (J+h)  \right)
  -\frac{1}{4}  \ln n - \frac{1}{8}  \ln \left(  \frac{1}{2J} (J-h)
  \right)
  + \frac{1}{12} \ln 2 + 3 \zeta^{\prime}(-1) + o(1).
\end{equation}
%\end{widetext}
(Here $\zeta(x)$ denotes the Riemann zeta function.)
Other asymptotic formulae were derived \cite{EFIK95b} for $h$ values
close to the saturation fields $h=\pm J$, where the ground state of
the XX chain becomes a completely polarized ferromagnet. A related $T=0$ correlation function
involving a weakly ferromagnetic string of spins was calculated
 \cite{AK02} using bosonization methods, and the value of $P(6)$ was
calculated \cite{BKS02} for the $XXX$ model ($\Delta=1$) at $T=h=0$.
For arbitrary $|\Delta| \leq 1$,
an asymptotic formula (for $T=h=0$) of the form 
\begin{equation}
  \label{eq:1.2b}
  P(n) \approx A n^{-\gamma} C^{-n^2}
\end{equation}
with explicit expressions for $C$ and $\gamma$ (as functions of
$\Delta$) was suggested \cite{KLNS02}. This formula reproduces the known exact results for $\Delta=0,
\frac{1}{2}$ and compares favorably to numerical data for other
$\Delta$ values.

For nonzero temperature the EFP correlation of the XXZ chain is
expected to show exponential decay for sufficiently large $n$.
The following simple argument for this asymptotic form was given by
Boos and Korepin
\cite{BK01}.
The EFP can be defined as the 
expectation value of the projection operator onto a state with
$n$ adjacent up spins: 
\begin{equation}
  \label{eq:1.3}
  P(n)=\left\langle \prod_{j=1}^n P_j \right\rangle = 
\frac
{\mathrm{Tr} e^{-\frac{H}{kT}}  \prod_{j=1}^n P_j }
{\mathrm{Tr} e^{-\frac{H}{kT}} },
\end{equation}
with
\begin{equation}
  \label{eq:1.4}
  P_j=S_j^z+\frac{1}{2}.
\end{equation}
The partition function for an $N$-site chain is
\begin{equation}
  \label{eq:1.5}
  Z_N = {\mathrm{Tr} e^{-\frac{H}{kT}} } = e^{-\frac{Nf}{kT}}
\end{equation}
in the limit of large $N$, where $f$ is the free energy per site.
Note  that the saturated ferromagnetic state with all 
spins up is a zero-energy eigenstate of the Hamiltonian
(\ref{eq:1.1}). (Let us temporarily consider the case $h=0$ for simplicity.) The
product of projection operators in the numerator of $P(n)$ 
(\ref{eq:1.3}) thus projects onto states with $n$ up spins, for which
the corresponding $n$-site ``partial Hamiltonian'' can be replaced by
zero. In other words, neglecting boundary effects, the numerator of
$P(n)$ is nothing but the partition function of an $(N-n)$-site chain:
\begin{equation}
  \label{eq:1.6}
  P(n) \sim \frac{Z_{N-n}}{Z_N} \sim
  \frac{e^{-\frac{(N-n)f}{kT}}}{e^{-\frac{Nf}{kT}}} = e^\frac{nf}{kT}.
\end{equation}
 The neglect of boundary effects in the
above ``derivation'' becomes increasingly problematic at low $T$,
because the zero-temperature correlations in the model (\ref{eq:1.1})
have long-range power law decays.

A finite magnetic field $h$ does not change the above argument essentially;
of course the free energy per particle depends on $h$. The
ferromagnetic state still is an eigenstate of the Hamiltonian, but its
energy eigenvalue changes from zero to $-h/2$ per site; accordingly
the asymptotic form of $P(n)$ is
\begin{equation}
  \label{eq:1.7}
  P(n) \sim \exp \left(  \frac{n}{kT} \left( f(T,h)+\frac{h}{2}\right)  \right).
\end{equation}
It should be noted that nowhere in the above line of argument we
exploited the fact that the nearest-neighbor exchange interactions $J$
in the Hamiltonian (\ref{eq:1.1}) are all equal; the asymptotic formula
(\ref{eq:1.7}) thus should hold also for modulated (for example
alternating) exchange interactions. However, for modulations with very long
wavelength or for random couplings the concept of a constant free
energy per site loses its meaning.

In the XXZ chain (\ref{eq:1.1}) the total spin $z$ component 
$S_{\mathrm{tot}}^z = \sum_{i=1}^N S_i^z$ is a conserved
quantity. That is different for the (anisotropic) XY chain
\cite{LSM61,BM71}
\begin{equation}
  \label{eq:1.8}
  H=J  \sum_{i=1}^N 
\left(
\vphantom{\frac {1}{2}} 
(1+\gamma) S_i^x  S_{i+1}^x +(1-\gamma) S_i^y  S_{i+1}^y
\right) 
-h  \sum_{i=1}^N  S_i^z
\end{equation}
with anisotropy parameter $\gamma$. The %ground-state phase diagram the
$(\gamma,h)$-plane shows several ground-state phases. The XX model discussed above
corresponds to the line $\gamma=0$, where the ground-state EFP is
asymptotically Gaussian \cite{STN01}. Away from that line, however,
the EFP is asymptotically exponential \cite{AF03,FA05,Fra08}. For
nonzero temperature the EFP is asymptotically exponential for
arbitrary $\gamma$ and $h$ \cite{FA05}.

A model which shows surprising similarities, but also important differences to
the anisotropic XY chain (\ref{eq:1.8}) is the dimerized XX chain 
\begin{equation}
  \label{eq:1.9}
   H=J  \sum_{i=1}^N
\left(
1-(-1)^i \delta
\right)
\left(
\vphantom{\frac {1}{2}} 
 S_i^x  S_{i+1}^x + S_i^y  S_{i+1}^y
\right) 
-h  \sum_{i=1}^N  S_i^z.
\end{equation}
The parameter $\delta$ quantifies the degree of dimerization; for $\delta=1$
the system decomposes into decoupled spin dimers.
Ground-state properties (in particular, dynamic correlations) were studied in parallel for
both models, (\ref{eq:1.8}) and  (\ref{eq:1.9}), by Taylor and
M\"uller \cite{TM85}. Dynamic properties at finite temperatures were studied
in \cite{DKS02} for the dimerized model and in \cite{DVKB06} for the
anisotropic model 
(including an additional Dzyaloshinskii-Moriya interaction). Both
models map to noninteracting Fermi quasiparticles under the
Jordan-Wigner transformation; see next section for details. For
$\gamma=0$ in  (\ref{eq:1.8}) and $\delta=0$ in  (\ref{eq:1.9}) the
two models are identical, with a gapless single-particle energy spectrum for
the Jordan-Wigner fermions. For
small nonzero $\gamma$ or $\delta$ and moderate magnetic field
$h$ a gap appears in the single-particle spectra of both
models. However, the parameter dependence of the energy spectra
 and the physical
meaning of the quasiparticles are fundamentally different between the
two models. In the dimerized model, the number of quasiparticles is a
conserved quantity strictly related to the total spin $z$ component
$S_{\mathrm{tot}}$. In the anisotropic model, in contrast,
$S_{\mathrm{tot}}$ is no longer conserved and the Bogoliubov rotation
involved in the diagonalization of the model mixes creation and
annihilation operators (corresponding to spin raising and lowering
operators). Consequently  the ground state of this model is very
simple in terms of the Jordan-Wigner quasiparticles, while it is a
complicated superposition involving many $S_{\mathrm{tot}}$
eigenstates in terms of the original spins. The shape of the
single-particle spectrum in the dimerized model depends only on
$\delta$ ($J$ fixing the overall energy scale). The spectral gap
separates two bands containing equal numbers of single-particle
states. The magnetic field $h$ (equivalent to a chemical potential)
determines the occupations of the bands, or, in spin language, the
total magnetization  $S_{\mathrm{tot}}$. In the anisotropic model, in
contrast, the shape of the single-particle spectrum depends on both
$h$ and $\gamma$, the two energy bands contain different numbers of
states, the lower quasiparticle band is always filled, and the upper
band is always empty, see \cite{TM85} for more details about the
ground states of the two models. 

It is thus of interest to study the EFP correlation for the dimerized
model and to compare the results to those obtained
\cite{AF03,FA05,Fra08} for the anisotropic model. Our results show
that the EFP of the dimerized system is asymptotically Gaussian at
$T=0$ and exponential at $T>0$. The exponential behavior is described
by the simple free energy argument leading to (\ref{eq:1.7}).

The remainder of the paper is organized as follows. In section
\ref{II} we explain the numerical procedure which can be used to
calculate $P(n)$ for arbitrary inhomogeneous  XX chains. Section
\ref{III} contains numerical results for the homogeneous chain which
are compared to the expected asymptotic behavior. Finite-size and
boundary effects are also discussed. Section \ref{IV} presents
results for the dimerized chain, with two different values for the
dimerization parameter.

%%%%%%%%%%%%%%%%%%%%%%%%%%%%%%%%%%%%%%%
%%%%%%%%%%%%%%%%%%%%%%%%%%%%%%%%%%%%%%

\section{The numerical procedure}
\label{II}

We consider the general $S=1/2$ $XX$ chain 
\begin{equation} 
  \label{eq:2.1}
  H= \sum_{i=1}^{N-1} J_i (S_i^x S_{i+1}^x +S_i^y S_{i+1}^y) -h
  \sum_{i=1}^N S_i^z
\end{equation}
which is one of the simplest quantum many-body systems conceivable, because many of its
properties can be derived from those of noninteracting lattice
fermions.

The Hamiltonian
(\ref{eq:2.1})  describing an open-ended $N$-site spin-1/2 $XX$ chain  can be mapped to a Hamiltonian of noninteracting fermions,
\begin{equation}
  \label{eq:2.2}
  H_F=  \frac{1}{2} \sum_{i=1}^{N-1} J_i (c_i^{\dag} c_{i+1} +
  c_{i+1}^{\dag} c_i) 
- h \sum_{i=1}^N (c_i^{\dag} c_i - \frac{1}{2})
\end{equation}
by means of the Jordan-Wigner transformation \cite{LSM61,Kat62} between spin
and Fermi operators:
\begin{equation}
  \label{eq:2.3}
  P_i= S_i^z +\frac{1}{2} = c_i^{\dag} c_i ,
\end{equation}
\begin{equation}
  \label{eq:2.4}
  S_i^+ = (-1)^{\sum_{l=1}^{i-1}c_l^{\dag} c_l} c_i^{\dag} = \prod_{l=1}^{i-1}
  (1 - 2 c_l^{\dag} c_l) c_i^{\dag}.
\end{equation}
In the homogeneous case $J_i \equiv J$, the fermion Hamiltonian is
\begin{equation}
  \label{eq:2.5}
  H_F = \sum_k \varepsilon_k c_k^{\dag} c_k + \frac{Nh}{2}
\end{equation}
where the operators $c_k^{\dag}$ and $ c_k$ create and destroy a
fermion in a one-particle eigenstate, respectively. The
one-particle energy eigenvalues
are
\begin{equation}
  \label{eq:2.6}
  \varepsilon_k = J \cos k-h , k= \frac{\nu \pi}{N+1}
, \nu= 1, \cdots,N
\end{equation}
and the eigenvectors are sinusoidal functions of the site index $i$. 
For a dimerized chain, where $J_i$ alternates between even and odd $i$
(compare (\ref{eq:1.9})), the eigenvalues $\varepsilon_k$ and the
corresponding eigenvectors are known analytically \cite{TM85}. In the
notation of the Hamiltonian  (\ref{eq:1.9}) the one-particle energies
are given by 
\begin{equation}
  \label{eq:2.6b}
  \varepsilon_k = -h + J \mathrm{sign}(\cos k) \sqrt{ \cos^2 k + \delta^2
    \sin^2 k}.
\end{equation}
It should be noted that (\ref{eq:2.6b}) was derived in the thermodynamic limit $N \to \infty$. For finite $N$ it is important to distinguish between even and odd chain lengths in the dimerized case. Detailed discussions of both cases can be found in \cite{FR05,KF06}. The data presented in the present paper were obtained by numerical diagonalization without making explicit use of the dispersion relations (\ref{eq:2.6},\ref{eq:2.6b}).

The ground state has all single-particle states with negative energies
occupied by Jordan-Wigner fermions while all other states are
empty. For $|h| < J \delta$ the zero-energy level lies within the
spectral gap of (\ref{eq:2.6b}) and hence the ground state does not
depend on $h$ in this range. The same is true for $|h|>J$ where the
ground state is either completely occupied by Jordan-Wigner fermions
or completely empty. In the intermediate field range, $J\delta < |h| <
J$, the groound state contains a partially filled band of
Jordan-Wigner fermions. Consequently we may expect behavior similar to
that of the homogeneous chain for $|h|<J$.

For a trimerized system, that is, $J_{i+3}=J_i$
in (\ref{eq:2.1}) the eigenvalues and eigenvectors still are known
analytically \cite{Gar01}, whereas for periodically varying couplings
with a larger period $p$ they can in general only be obtained by
numerically solving a $p \times p$ eigenvalue problem; see, for
example \cite{LAG06}. 
For
general $J_i$ neither eigenvalues nor eigenvectors are available analytically,
however, both are easily obtained from the solution of a tridiagonal
eigenvalue problem with standard numerical procedures \cite{PTVF92}.

The asymptotic formula (\ref{eq:1.7}) for the EFP correlation may be
evaluated more explicitly. Taking into account (\ref{eq:2.5}) and the
standard formula for the free energy of free fermions we obtain
\begin{equation}
  \label{eq:2.7}
  P(n)= c(T,h) \exp \left( -\frac{n}{kT} \left( \frac{kT}{N} \sum_k \ln (1+ e^{-\beta
      \varepsilon_k}) - \frac h2  \right)  \right).
\end{equation}
This approximate asymptotic formula for the EFP correlation should
hold as long as the concept of a spatially 
(roughly) constant free energy per site makes sense, as discussed
above. We have introduced a temperature- and field-dependent prefactor
$c(T,h)$. For the homogeneous $XX$ chain, Shiroishi et
al. \cite{STN01} have derived an explicit expression for $c(T,h)$.

The exact numerical evaluation of the EFP correlation proceeds as
follows. Due to (\ref{eq:2.3}) 
\begin{equation}
  \label{eq:2.8}
  P(n) = \langle c_1^{\dag} c_1 c_2^{\dag} c_2 \cdots c_n^{\dag} c_n \rangle.
\end{equation}
This many-fermion expectation value can be evaluated using the
Wick-Bloch-De Dominicis Theorem \cite{Gau60}. Mathematically this means
that the expectation value (\ref{eq:2.8}) can be expressed as a
Pfaffian involving only elementary expectation values. Pfaffians are
close relatives of determinants and we refer the reader to the
literature for their properties \cite{GH64}. The numerical evaluation
of Pfaffians proceeds along similar lines as that of
determinants. Elements can be reduced to zero by operations which are
known to leave the value of the Pfaffian invariant. After production
of sufficiently many zero elements the evaluation of the Pfaffian
becomes trivial due to an expansion theorem. An implementation along
these lines was described by Derzhko and Krokhmalskii \cite{DK98}. We
use a similar algorithm here. A
recursive scheme for evaluating Pfaffians was used by Jia and
Chakravarty \cite{JC06}.

%%%%%%%%%%%%%%%%%%%%%%%%%%%%%%%%%%%%%%%%%%%%%%%%%%%%%%%%%%%%%%%%%%%%%%%%%
\section{Results for the homogeneous case}
\label{III}
\subsection{$T=0$}

\begin{figure}[h]
\centerline{\includegraphics[width=0.65\textwidth]
{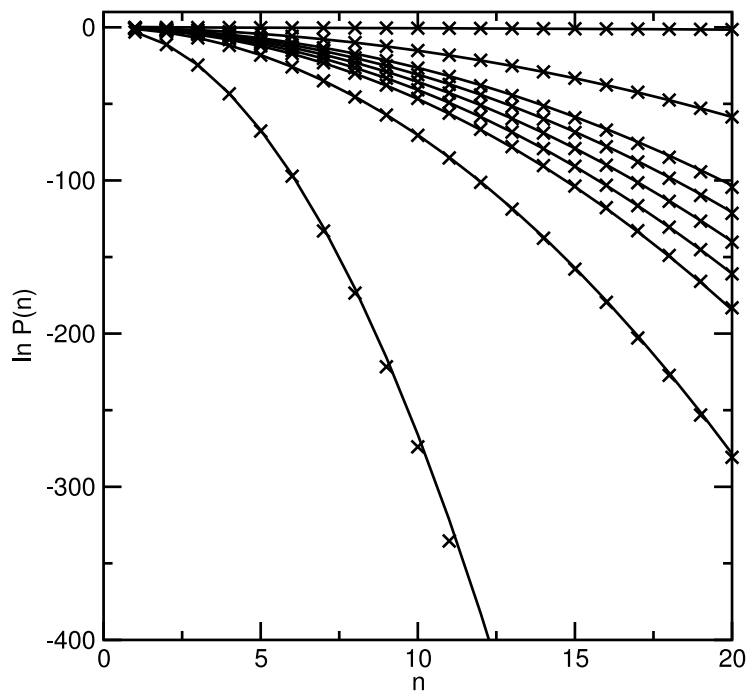}
}
\caption{The EFP correlation of the homogeneous  chain at $T=0$,
  for magnetic fields $h=0.99$, 0.5, 0.2, 0.1, 0, -0.1, -0.2, -0.5,
  -0.99 (top to bottom). Crosses represent the numerical evaluation,
  solid lines show the asymptotic behavior  (\ref{eq:1.2}).}
 \label{eins}
 \end{figure}

We have tested our numerical evaluation for chains with
$N=256$. Figure \ref{eins} shows the EFP correlation $P(n)$ for a
string of spins starting at site $i=110$, for several values of the
magnetic field $h$. 

In order to assess the influence of finite-size
and boundary effects we have also performed computations for $i=111$,
and also for $N=512$ and $i=221$. We found that at $T=0$ there are
still finite-size and even/odd effects of the order of up to one
percent. Given the fact that ground-state correlations in quantum spin
chains tend to be long-ranged, this does not come as a 
surprise. 

On the scale of the figure there is near perfect agreement
between numerical evaluation of the correlation and the asymptotic
formula (\ref{eq:1.2}) as already noted by Shiroishi et
al. \cite{STN01}. Only at extremely small values of the correlation
function (outside the range of the figure) 
there are significant deviations from the asymptotic
behavior. These are, however due to precision problems in the
numerical evaluation of  $P(n)$.

\subsection{$T>0$}

\begin{figure}[h]
\centerline{\includegraphics[width=0.65\textwidth]
{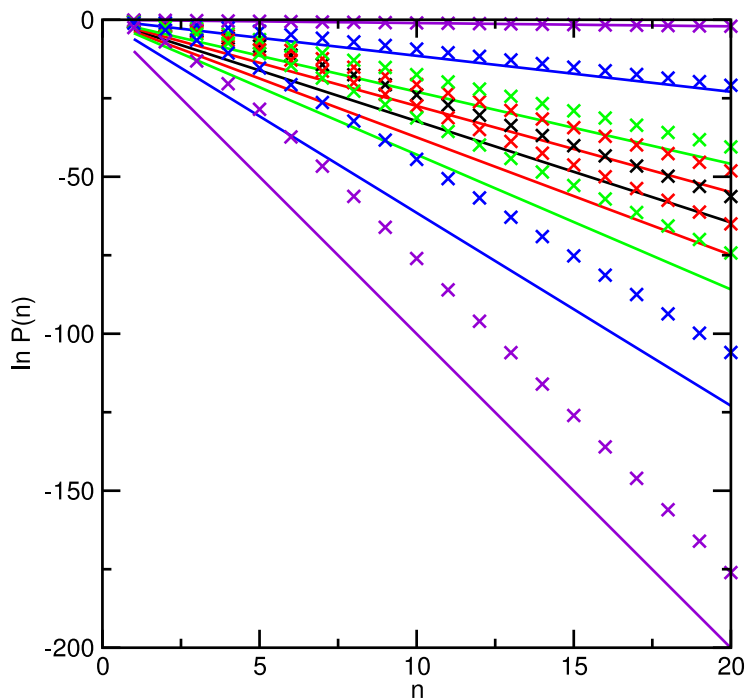}
}
\caption{The EFP correlation of the homogeneous  chain at $T=0.1$,
  for magnetic fields $h=0.99$, 0.5, 0.2, 0.1, 0, -0.1, -0.2, -0.5,
  -0.99 (top to bottom). Crosses represent the numerical evaluation,
  solid lines show the asymptotic behavior  (\ref{eq:2.7}), where we
  have put $c(T,h)=1$.}
 \label{zwei}
 \end{figure}

\begin{figure}
\centerline{\includegraphics[width=0.65\textwidth]
{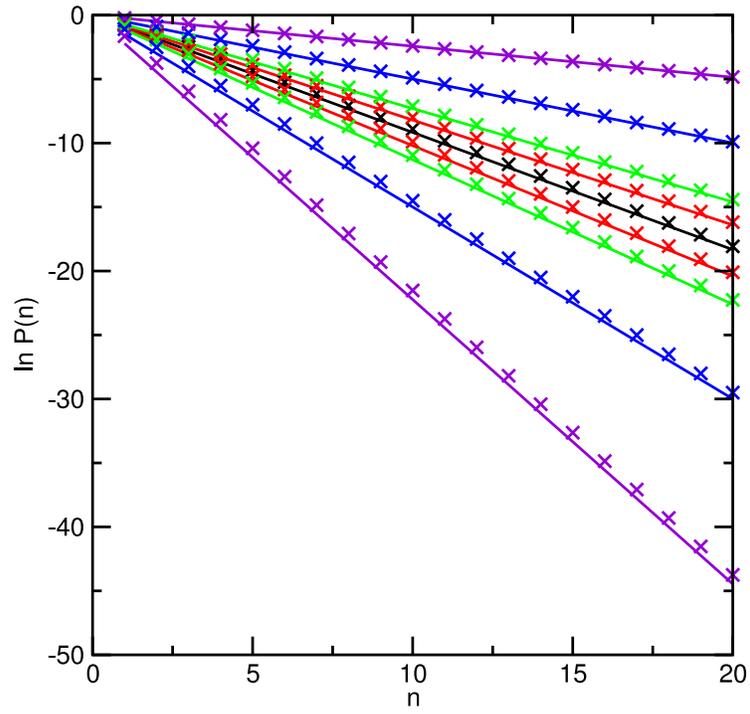}
}
\caption{Same as Figure \ref{zwei}, for $T=0.5$.}
 \label{drei}
 \end{figure}

\begin{figure}
\centerline{\includegraphics[width=0.65\textwidth]
{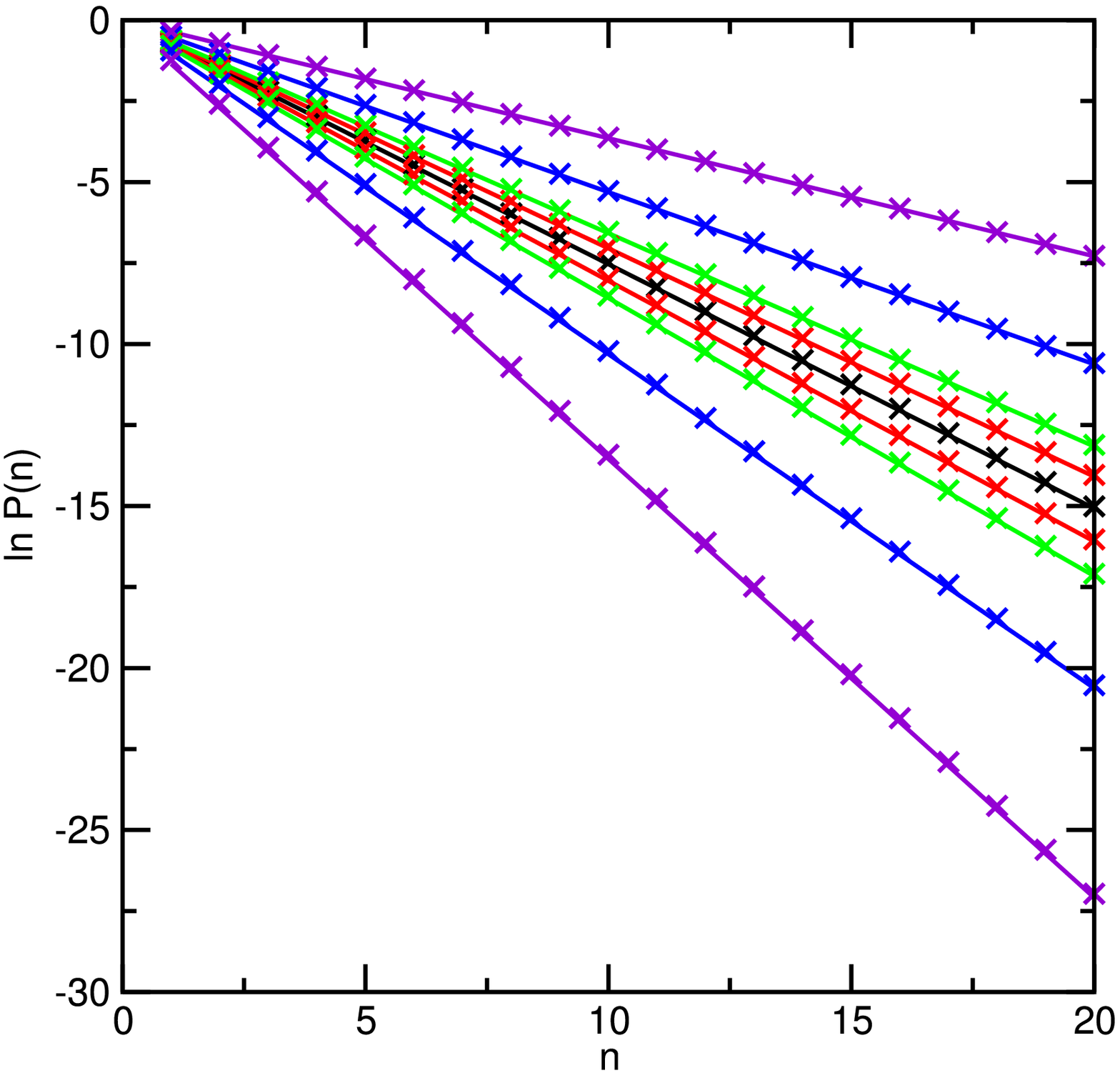}
}
\caption{Same as Figure \ref{zwei}, for $T=1$.}
 \label{vier}
 \end{figure}

\begin{figure}
\centerline{\includegraphics[width=0.65\textwidth]
{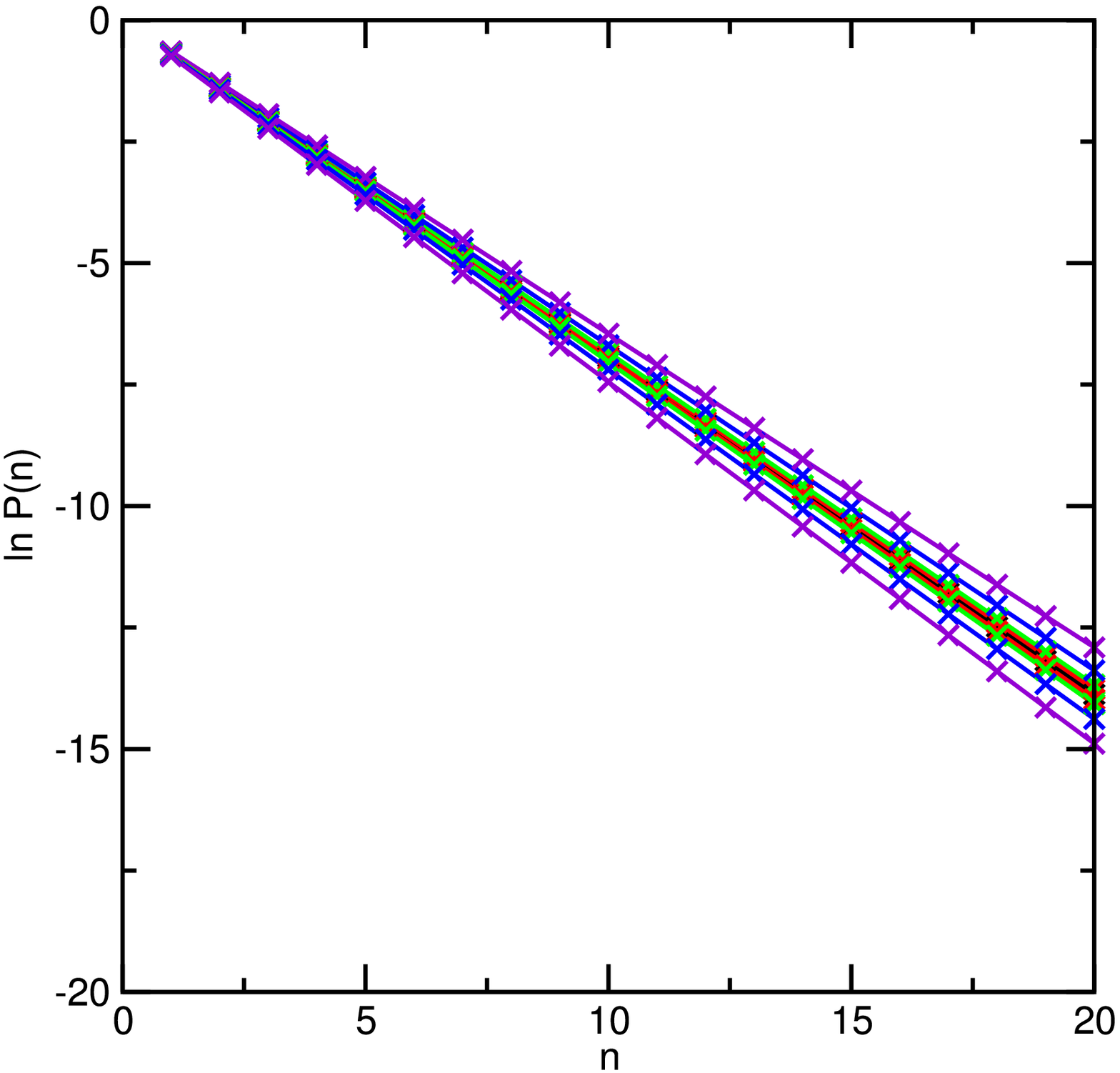}
}
\caption{Same as Figure \ref{zwei}, for $T=10$.}
 \label{fuenf}
 \end{figure}

\begin{figure}
\centerline{\includegraphics[width=0.65\textwidth]
{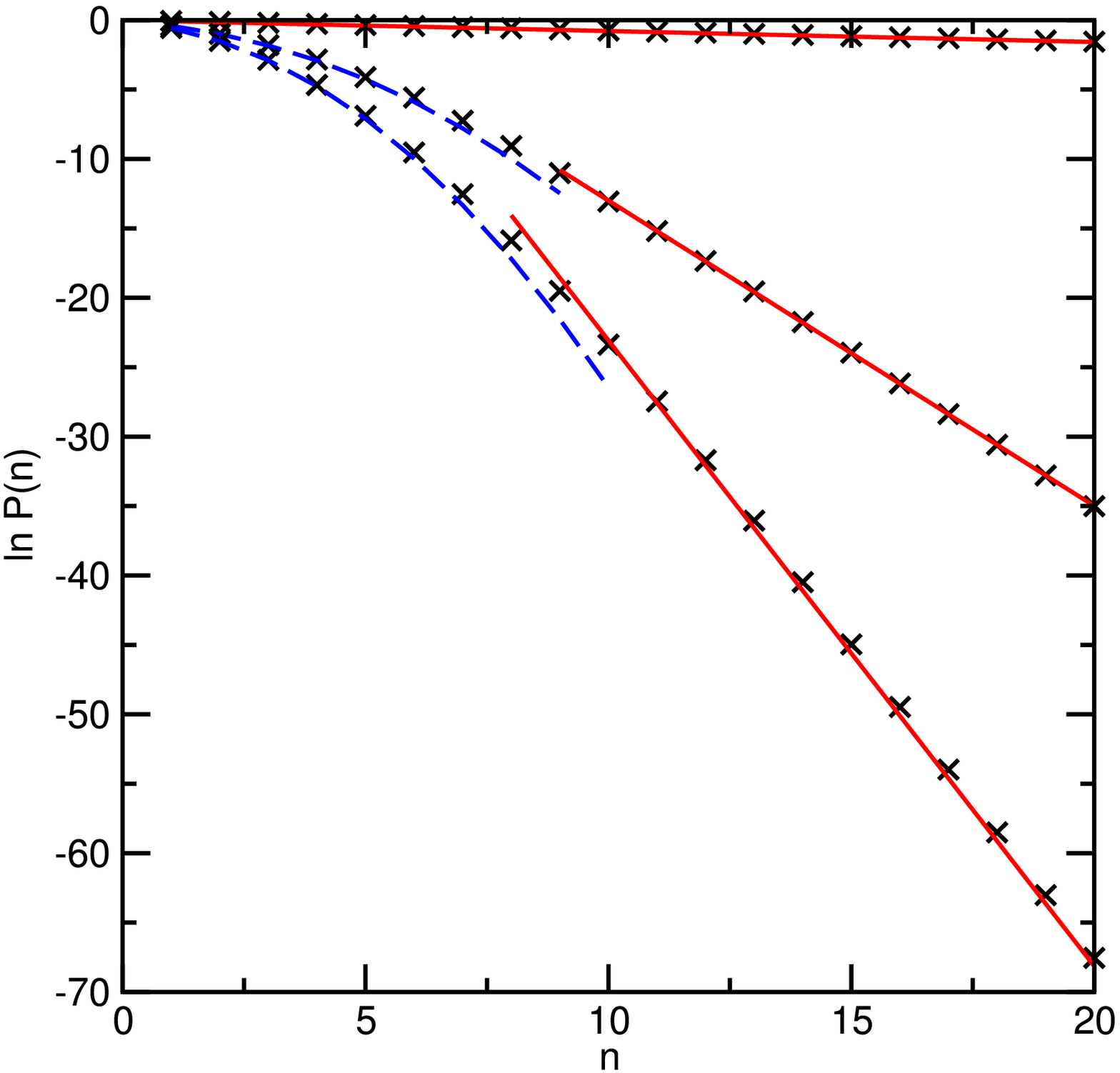}
}
\caption{The EFP correlation of the homogeneous  chain at $T=0.05$,
  for magnetic fields $h=0.99$, 0.5, and 0.2 (top to bottom). 
Crosses represent the numerical evaluation,
  solid (red) lines show the exponential asymptotic behavior
  (\ref{eq:2.7}), with adjusted value for
 $c(T,h)$, dashed (blue) lines show the Gaussian asymptotic behavior
 (\ref{eq:1.2}) for $h=0.5$ and 0.2.}
 \label{fuenf_b}
 \end{figure}

We have numerically evaluated the EFP correlation $P(n)$ for $N=256$
chains for a string of spins starting at site $i=90$ for several
values of $T$ and the magnetic field $h$. Figures
\ref{zwei},\ref{drei},\ref{vier}, and \ref{fuenf} show results for
$T=0.1$, 0.5, 1, and 10, respectively. 

In all cases it was found that
the asymptotic formula (\ref{eq:2.7}) describes the numerical data
very well. The prefactor $c(T,h) (\geq 1)$ is large at low $T$ and
negative values of $h$, as already observed by Shiroishi et
al. \cite{STN01} (compare their Figure 6). For intermediate and high
temperatures we observed a very small but seemingly systematic slope
when we plotted the difference between the logarithm of the numerical
value of $P(n)$ and the expression in the exponent of (\ref{eq:2.7})
versus $n$. Due to the smallness of the effect we did not further
pursue the origin of this difference, however. 

We have also assessed
the influence of finite-size and boundary effects on $P(n)$ for
$T=0.1$ and $h=\pm 0.99, \pm 0.1$, and 0. For these parameter values
we compared the numerical results for $N=256$, starting sites $i=90$
and 91, and $N=512$, $i=240$. No differences beyond the $10^{-7}$
level were found. 

In the limit $T \to \infty$
$P(n)=\left(\frac{1}{2}\right)^n$ for arbitrary $h$. That is precisely
what the $h=0$ results already show at $T=10$ (see Fig. \ref{fuenf}).
For some selected $h$ and $T$ values we have also compared our values
for the decay length of $P(n)$ to those given in Figure 5 of Shiroishi
et al. \cite{STN01}. The results agree.

At low temperature a crossover from the Gaussian $T=0$ behavior (\ref{eq:1.2}) at
small $n$ to the exponential asymptotics (\ref{eq:2.7}) at larger $n$ can be
observed, as shown in Fig. \ref{fuenf_b}.

\clearpage

\section{Results for the dimerized case}
\label{IV}

We have calculated $P(n)$ for weak and strong dimerization,
$\delta=0.2$ and $\delta=0.8$.

\subsection{$\delta=0.2, T=0$}

For given dimerization $\delta >0$ there are two critical values for
the magnetic field, $h=\pm \delta$. For $|h|<\delta$ the ground state
does not depend on $h$. In the Fermion picture it consists of a
completely filled lower band and a completely empty upper band. It may
thus be expected that correlation functions do not change with $h$ for
$h$ within this range. 

\begin{figure}
\centerline{\includegraphics[width=0.65\textwidth]
{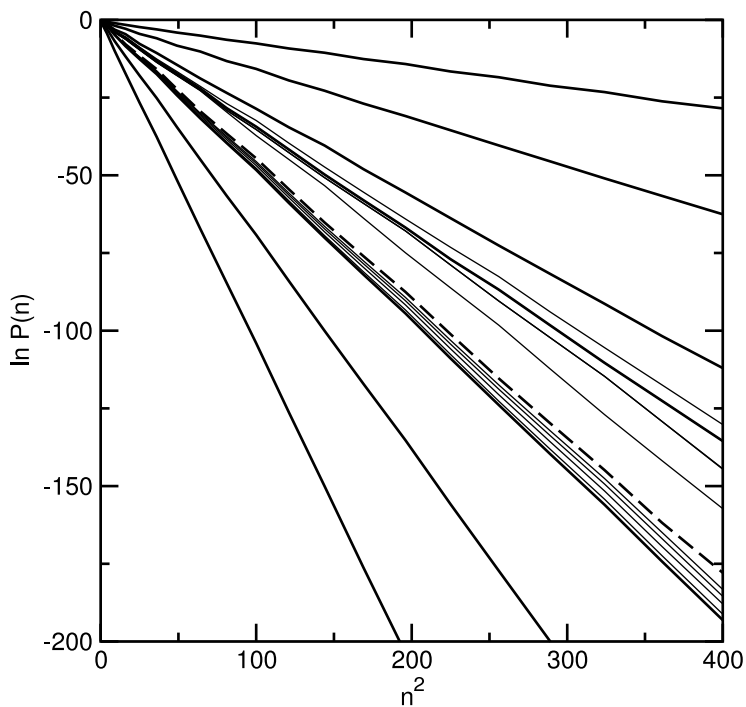}
}
\caption{The EFP correlation $P(n)$ of the dimerized chain as a function of
  $n^2$ for $T=0, \delta=0.2$ and $h=-0.75$, -0.5, -0.25, -0.24,
  -0.23, -0.22, -0.21, -0.20, 
 +0.201, 0.202, 0.203, 0.204,
  0.205, 0.21, 0.25, 0.5, and 0.75 (bottom to top). 
Heavy lines denote the data for $h=-0.75$, -0.5, -0.25, -0.20 (dashed), +0.205, 0.25, 0.5, and 0.75. On the scale of this figure the data for $h=-0.20, 0,$ and +0.20 coincide. 
Note the rapid
  variation for $h$ slightly larger than 0.20. (The data for $h=0.202$ and 0.203 coincide on this scale, as do the data for $h=0.204$ and 0.205.)  }
 \label{sechs}
 \end{figure}

\begin{figure}
\centerline{\includegraphics[width=0.65\textwidth]
{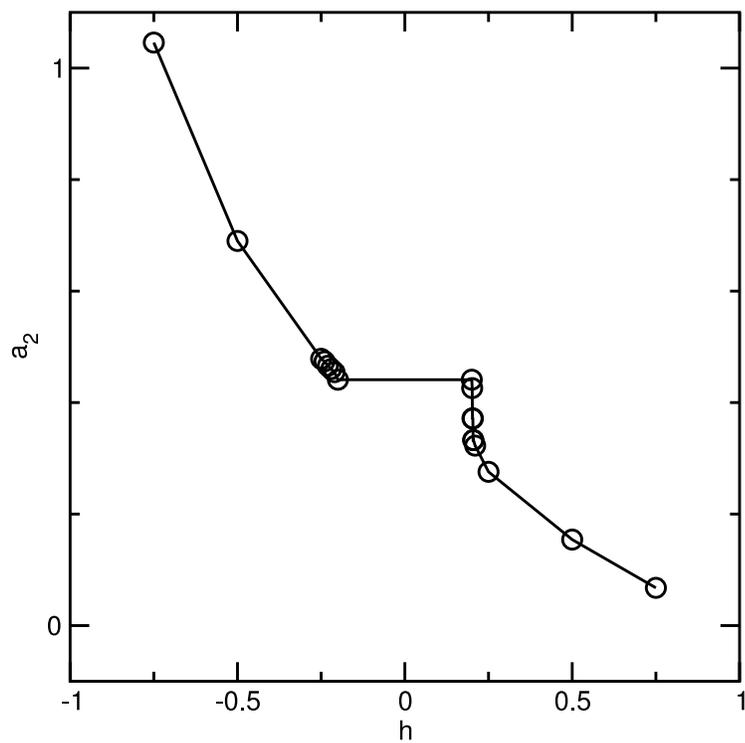}
}
\caption{Gaussian decay parameter $a_2$ obtained from the numerical
  $P(n)$ data shown in Fig. \ref{sechs}.} 
 \label{acht}
\end{figure}

\begin{figure}
\centerline{\includegraphics[width=0.65\textwidth]
{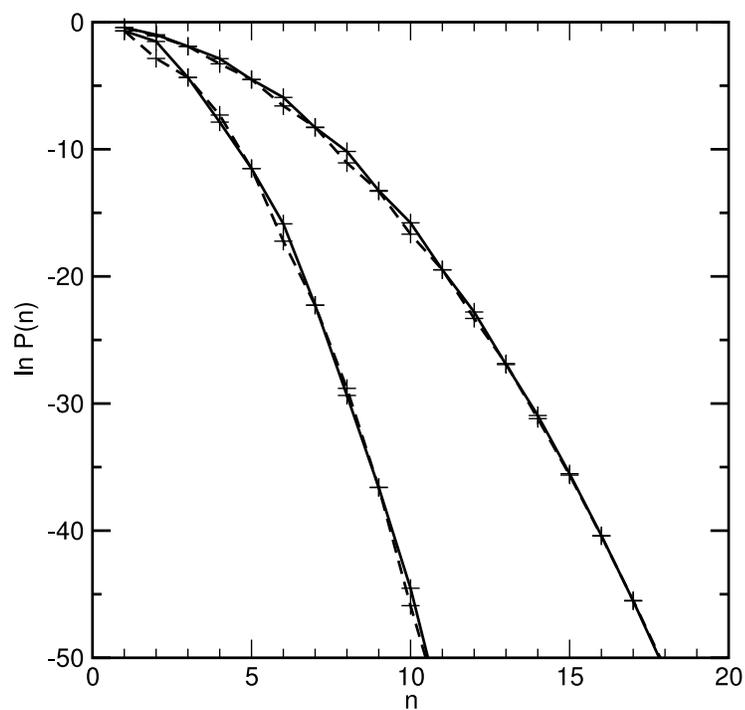}
}
\caption{The EFP correlation $P(n)$  of the dimerized chain  for $T=0$, $\delta=0.2$, $h=0.5$
  (upper set of curves) and $h=0.2$ (lower set of curves). The strings
  of sites considered start at site $i=110$ (solid lines) or $i=111$
  (dashed lines) of a $N=256$ chain.}
 \label{neun}
\end{figure}

Figure \ref{sechs} shows $P(n)$ for a large number of $h$ values, with
special attention to the critical regions near $h=\pm \delta$. 
The plot of $\ln P(n)$ vs $n^2$
convincingly demonstrates an overall Gaussian behavior, but a closer look
at the region $n \leq 10$, for example, reveals some
oscillations due to the dimerization. As expected, the EFP correlation
does not change as $h$ varies between $-\delta$ and $\delta$.

Figure \ref{acht} shows the values of
the Gaussian decay parameter $a_2$ obtained from the data of Figure
\ref{sechs} by a least-squares fit of
the form $\ln P(n) = a_0 + a_1 n - a_2 n^2$. The lines between the
data points are guides to the eye only. The band-edge singlarity at
$h=\delta$ is very prominently visible. In contrast, the behavior near
$h=-\delta$, where the lower quasiparticle band approaches complete
filling, is less singular.

The changes caused by varying the initial site $i$ are shown in Figure
\ref{neun} for a sub-critical field $h=0.2$ and a super-critical field
$h=0.5$. The solid and dashed lines are for $i=110$ and 111,
respectively. The $P(n)$ values for even and odd $i$ seem to coincide
for odd string lengths $n$. A closer look  at the data shows that at
$h=0.2$ this coincidence of data for odd $n$ is indeed perfect.
For
$h=0.5$ differences at the 1\% level show up for all odd $n$. This
behavior is not surprising since for odd $n$ the string of $n$ sites 
contains equal numbers of ``weak'' and ``strong'' bonds for both even
and odd $i$, hence the EFP as a global quantity of the whole string
should be the same, at least in the limit of infinite system size. For
even $n$, however, differences are to be expected, and are indeed
present in the numerical data. As an extreme
example, consider $n=2$: $P(2)$ refers to a single bond and will
significantly depend on whether that bond is strong or weak. 

In order to assess the effects of system size we have also compared
the $N=256$ data to $N=512$ data (for $i=220$ and 221). The data for
$h=0.2$ do not  depend on $N$, whereas the
$h=0.5$ data vary at the percent level between the two $N$ values.

As we are dealing with finite open systems, boundary effects are an issue
to be considered. For a given absolute value of the dimerization parameter
$|\delta|$ the chain may either start with a ``strong'' bond or with a
``weak'' bond. It turns out that switching back and forth between these
two possibilities (at $N=256$) is equivalent to switching between even and odd
$i$- The equivalence is in
fact perfect for
subcritical values of $|h|$, for example $h=0$ and 0.2. At $h=\pm0.5$
we observe , differences at the percent level, the size of which does
not change significantly between $N=256$ and $N=512$.

It should be an interesting task to generalize the asymptotic formula
(\ref{eq:1.2}) to the dimerized case along the lines of \cite{STN01} 
in order to better
understand the behavior observed in Fig. \ref{acht}.

\subsection{$\delta=0.2, T > 0$}

\begin{figure}
\centerline{\includegraphics[width=0.65\textwidth]
{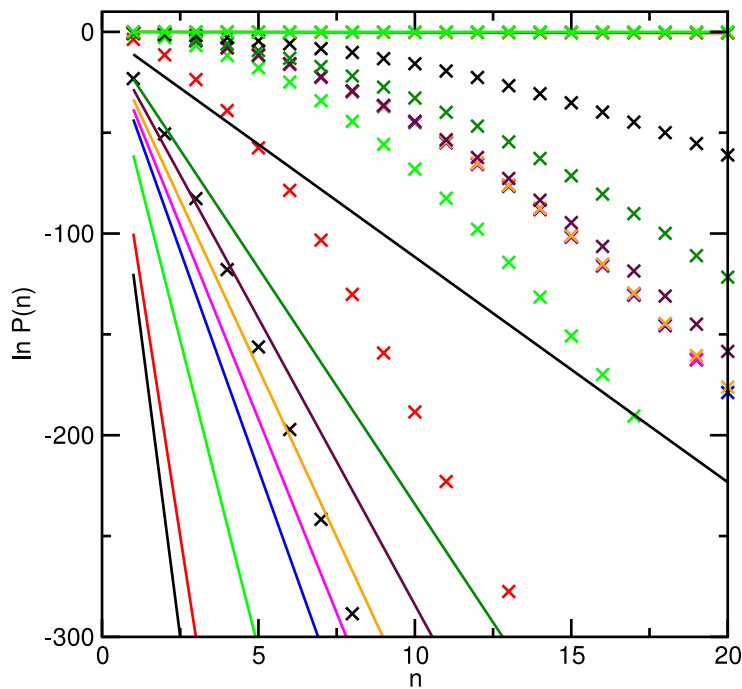}
}
\caption{The EFP correlation $P(n)$ of the dimerized chain   for $T=0.01$, $\delta=0.2$, and
  $h=1.2$, 1, 0.5, 0.2, 0.1, 0, -0.1, -0.2, -0.5, 1-, -1.2 (top to
  bottom). Symbols correspond to numerically calculated values, solid
  lines to the asymptotic formula \ref{eq:2.7}. On the scale of the
  figure the data for $h=1$ and 1.2 cannot be distinguished.}
 \label{zehn}
\end{figure}

\begin{figure}
\centerline{\includegraphics[width=0.65\textwidth]
{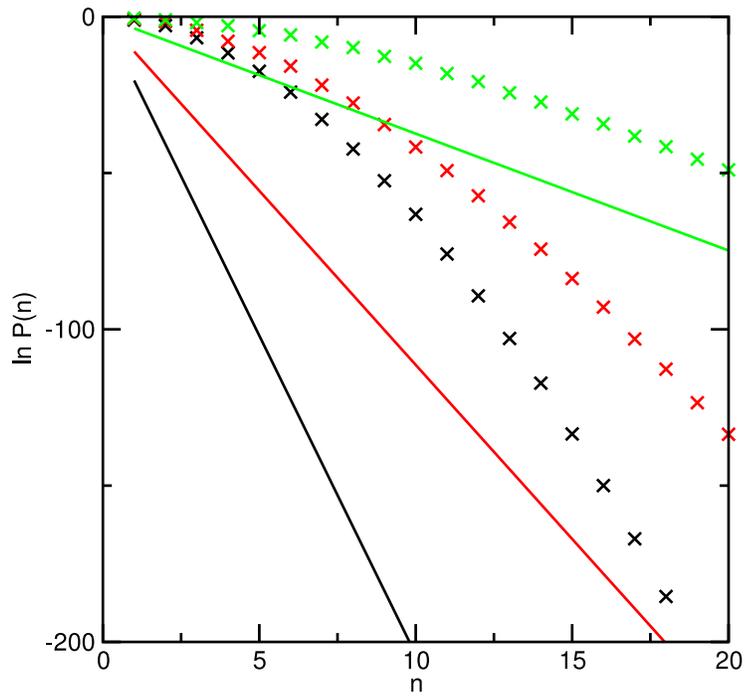}
}
\caption{The EFP correlation $P(n)$ of the dimerized chain   for $T=0.03$, $\delta=0.2$, and
  $h=0.5$, 0, and -0.5 (top to
  bottom). Symbols correspond to numerically calculated values, solid
  lines to the asymptotic formula \ref{eq:2.7}.}
 \label{elf}
\end{figure}

\begin{figure}
\centerline{\includegraphics[width=0.65\textwidth]
{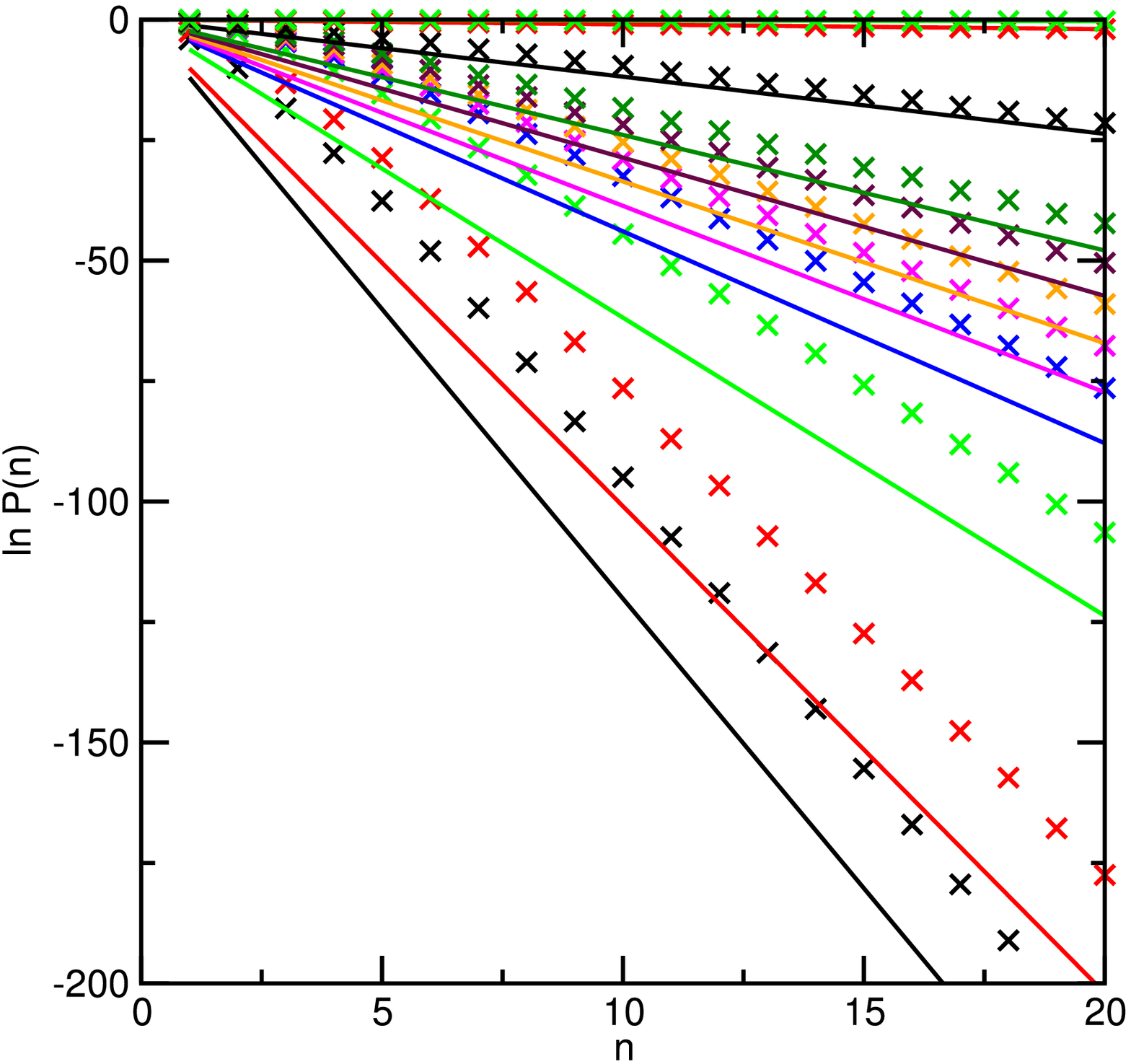}
}
\caption{Similar to Fig. \ref{zehn}, for $T=0.1$}
 \label{zwoelf}
\end{figure}

\begin{figure}
\centerline{\includegraphics[width=0.65\textwidth]
{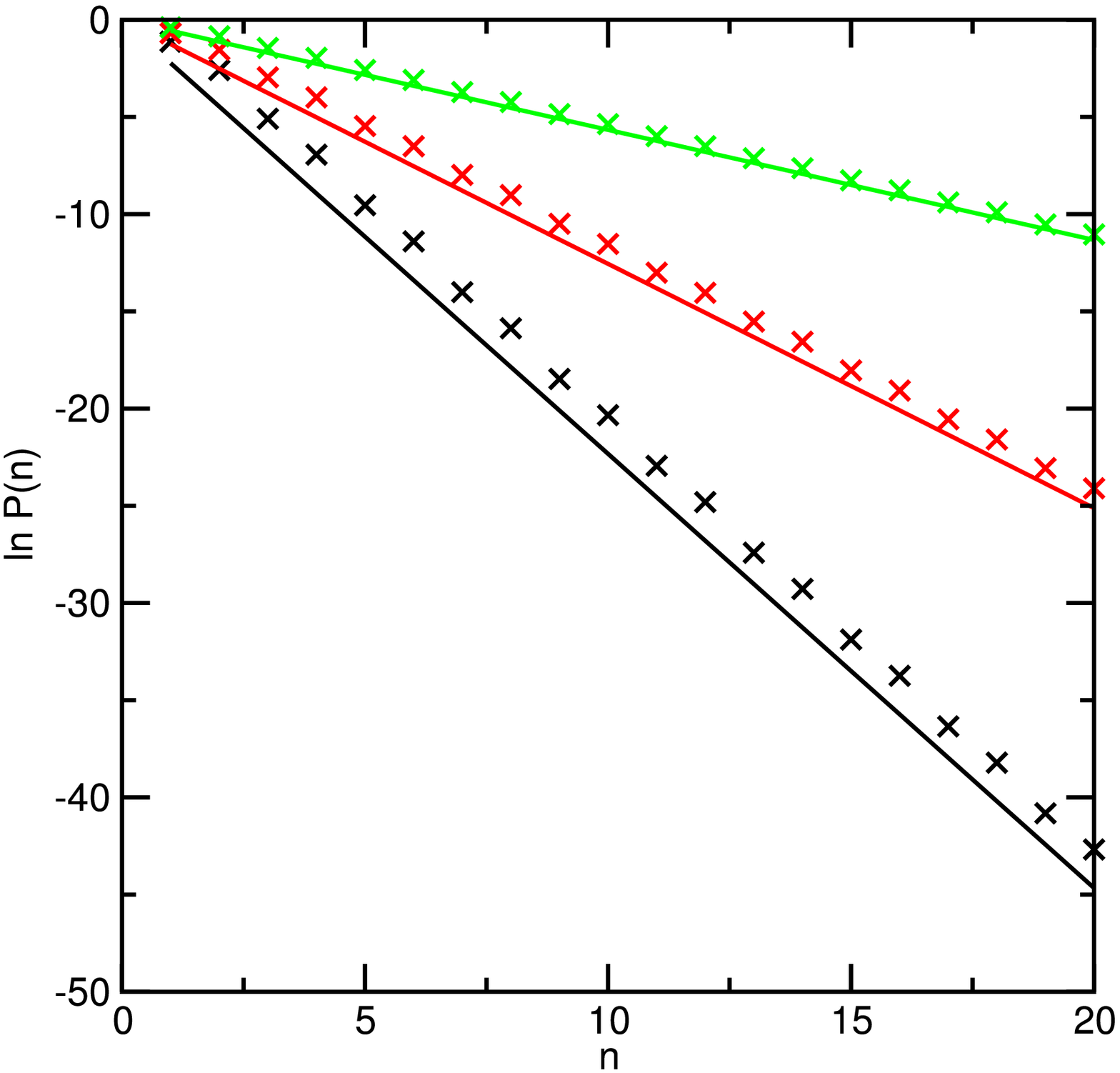}
}
\caption{Similar to Fig. \ref{elf}, for $T=0.3$}
 \label{dreizehn}
\end{figure}

Figure \ref{zehn} shows data for the low temperature $T=0.01$ and a
wide range of $h$ values. The strings of sites considered again start
at site $i=110$ of a $N=256$ chain. Comparing the numerically
calculated values of $P(n)$ to the asymptotic formula (\ref{eq:2.7})
we see clearly that at this low temperature the EFP correlation for $n
\leq 20$ still displays the typical  Gaussian
zero-temperature behavior. The correlations for $h=-0.2$, -0.1, and 0
stay  together very closely for $n \leq 20$, whereas those for
$h=0.1$ start to deviate for $n>10$ and those for $h=0.2$ even
earlier. This is in contrast to $T=0$ (see Fig. \ref{sechs}) where the
correlations for all $|h| \leq 0.2$ coincide because the ground state
does not change in this $h$ range.

This behavior changes as $T$ increases, as shown in Figures \ref{elf}
, \ref{zwoelf}, and \ref{dreizehn}, for $T=0.03$, 0.1, and 0.3,
respectively. Figure \ref{elf} shows a crossover of $P(n)$ from
Gaussian behavior at small $n$ to exponential behavior at larger $n$.
However, at this rather low $T$ value the prefactor $c(T,h)$ in the
asymptotic formula (\ref{eq:2.7}) is still quite large. Superimposed
on the asymptotic behavior the numerical data at $T=0.03$ and $h=0.5$
show the modulation expected due to the alternating exchange
couplings. That modulation becomes more visible and the prefactor
$c(T,h)$ approaches unity as $T$ grows, see Figs. \ref{zwoelf} and
\ref{dreizehn}. 

\subsection{$\delta=0.8$}

\begin{figure}
\centerline{\includegraphics[width=0.65\textwidth]
{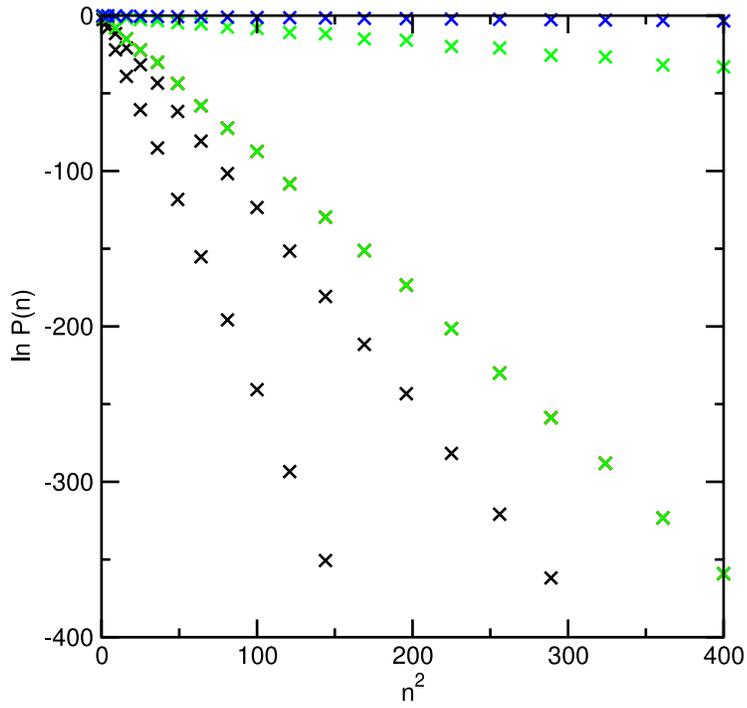}
}
\caption{The EFP correlation $P(n)$ of the dimerized chain   for $T=0$, $\delta=0.8$, and
  $h=0.99$, 0.9, 0.8, 0, -0.8, -0.9 and -0.99 (top to
  bottom). The data for $h=0, \pm0.8$ coincide.}
 \label{fuenfzehn}
\end{figure}

\begin{figure}
\centerline{\includegraphics[width=0.65\textwidth]
{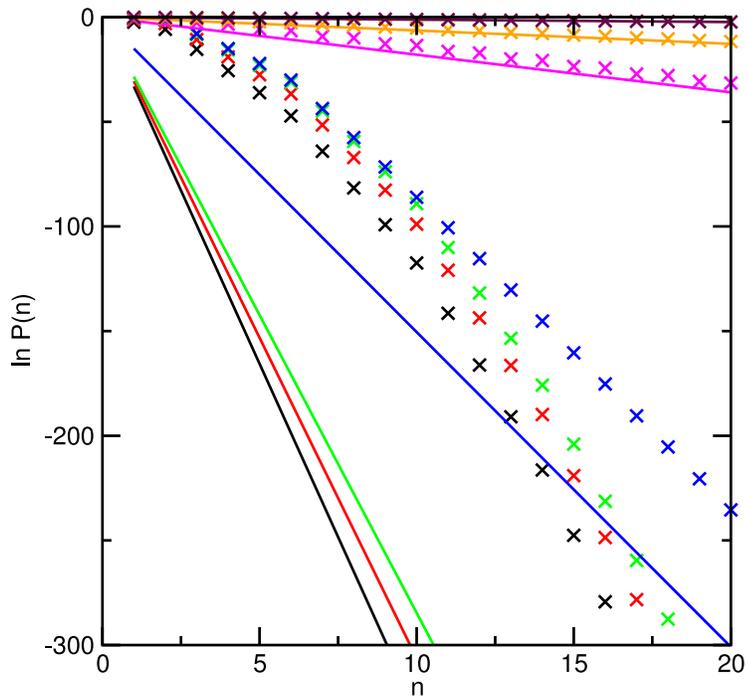}
}
\caption{The EFP correlation $P(n)$ of the dimerized chain   for $T=0.03$, $\delta=0.8$, and  $h=0.99$, 0.9, 0.8, 0, -0.8, -0.9 and -0.99 (top to
  bottom).  Symbols denote numerically calculated values, solid lines show
  the asymptotic formula  (\ref{eq:2.7}), with $c(T,h)=1$.}
 \label{sechzehn}
\end{figure}

\begin{figure}
\centerline{\includegraphics[width=0.65\textwidth]
{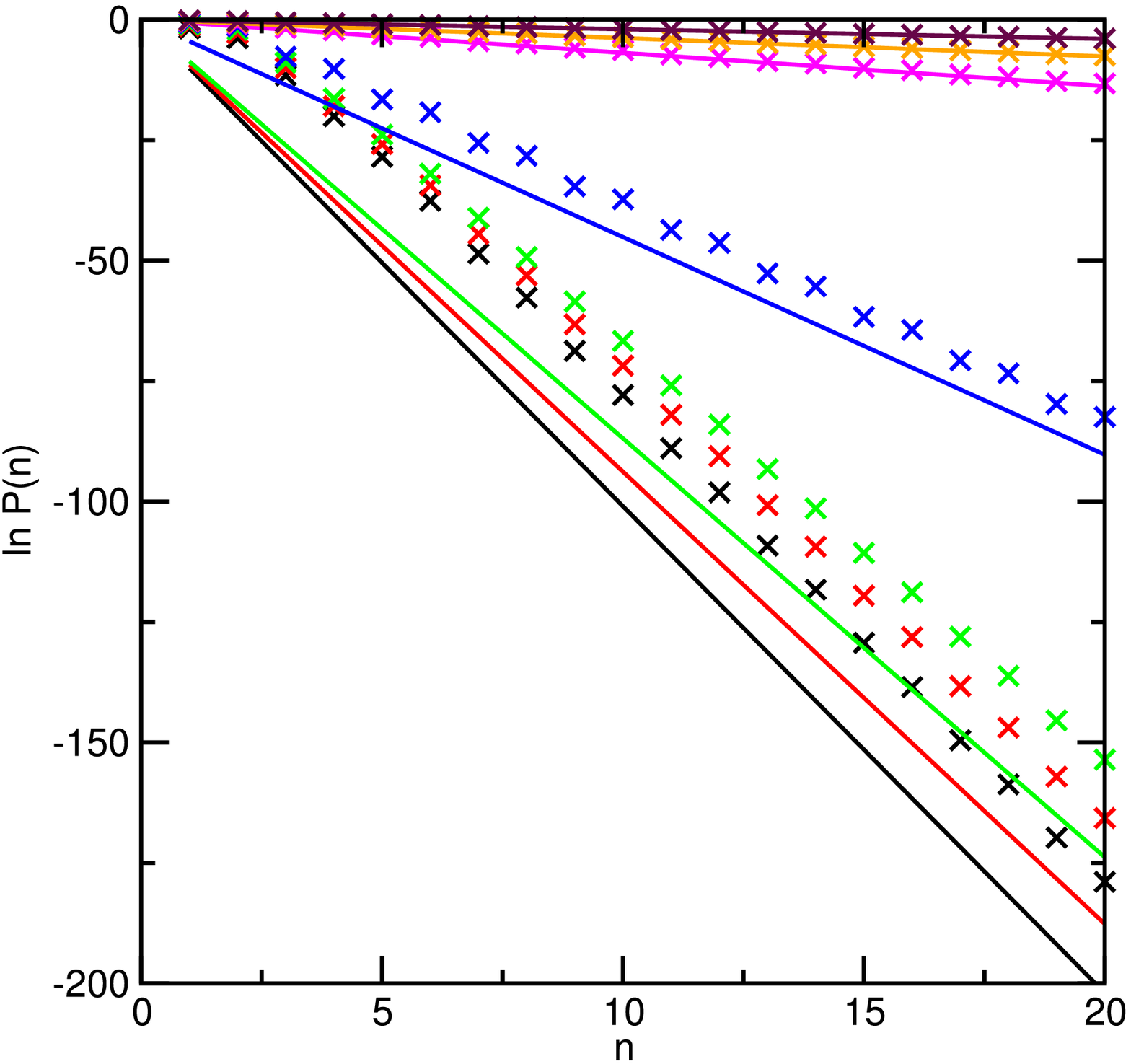}
}
\caption{Same data as Figure \ref{sechzehn}, for $T=0.1$.}
 \label{siebzehn}
\end{figure}

For strong dimerization we present data for $h=\pm0.99, \pm0.9,
\pm0.8$, and 0.  Figure \ref{fuenfzehn} shows the results for
$T=0$. 
The plot of
$\ln P(n)$ as a function of $n^2$ shows
that the decay is still Gaussian apart from the expected modulation
due to the dimerization, which for $\delta=0.8$ is of course expected
to be more visible than for $\delta=0.2$. Note that the data points
for $h=0, \pm0.8$ 
coincide as the ground state stays the same for thes $h$ values.

At $T=0.03$ (Figure \ref{sechzehn}) the data seem to fall naturally into three
groups, namely $h<0$, $h=0$, and $h>0$. The $h<0$ data do not seem to
reach the proposed asymptotic behavior (\ref{eq:2.7}) within the range $n \leq 20$,
whereas the other data do so quite nicely.  This is not true, however,
as closer inspection of the numbers reveals. The data for  $h<0$ reach
the asymptotics later and with a much larger prefactor $c(T,h)$ in  (\ref{eq:2.7}).

At $T=0.1$ (Figure
\ref{siebzehn}) all $n \geq5$ data fit to the expected exponential behavior
apart from the expected modulation due to dimerization. Note that at this `high'
temperature the $h=0$ data stand out alone and the data for the
critical field values $h=\pm0.8$ are much closer to the data for
$h=\pm0.9$ and $\pm0.99$, in contrast to the situation at $T=0$
(Fig. \ref{fuenfzehn}).

\section{Concluding Remarks}
\label{V}

We have examined the emptiness formation probability (EFP) correlation
for homogeneous and dimerized spin-1/2 XX chains numerically by
evaluating Pfaffian forms. For the homogeneous XX system the EFP is
known to show asymptotically Gaussian behavior in the ground state and
exponential behavior at finite temperature. For a homogeneous
(anisotropic) XY chain the EFP is known to be exponential at both zero
and finite temperatures. The anisotropic XY chain has a gapped
single-particle energy spectrum, as does the dimerized XX
chain. However, the EFP correlations of the two chains behave
differently. Our calculations show that the EFP of the dimerized chain
 is Gaussian at at zero
temperature and exponential at finite temperature. The exponential
behavior at finite temperature can be interpreted within a simple free
energy picture developed for the homogeneous chain. The generalization
of the $T=0$ asymptotic expansion from the homogeneous to the dimerized
system is left as a task for future work.

\section*{Acknowledgments}
JS is grateful to Masahiro Shiroishi for helpful discussions and correspondence.

% Create the reference section using BibTeX:

\bibliography{jsbas_def,general}

\end{document}